\documentclass[aps,prc]{revtex4}

\usepackage{amsmath}
\usepackage{amssymb,graphicx,array}
\usepackage[hypertex]{hyperref}

\begin{document}


\title{Neutrino emissivity under neutral kaon condensation}

\author{Sebastian Kubis}
\email{sebastian.kubis@ifj.edu.pl}
\affiliation{The Henryk Niewodnicza\'nski Institute of Nuclear Physics,\\
Polish Academy of Sciences,\\
ul.Radzikowskiego 152, 31-342 Krak\'ow, Poland}
\date{\today}

\begin{abstract}
Neutrino emissivity from neutron star matter  with neutral kaon condensate 
is considered. It is shown that a new cooling channel is opened, and
what is more, all previously known channels acquire greater emissivity 
reaching the level of the direct URCA cycle in normal matter.  
\end{abstract}


\maketitle

\def\ep{\varepsilon}
\def\pa{\partial}
\def\th{\theta}
\def\ni{\noindent}
\def\Tr{{\rm Tr}}
\def\fm3{\rm ~fm^3}
\def\MeV{\rm ~MeV}
\def\K0{\bar{K}^0}
\def\k0{\bar{k}^0}
\def\ee{{\rm e}}
\def\sp{s^\prime}
\def\anu{\bar{\nu}{}}
\def\dga{\Delta g\,}
\def\coschh{\cos{\chi\over 2}}
\def\gammc#1#2{\bar{#1}\,\Gamma^\mu_{#1#2}\, #2}
\def\gamms#1#2{\bar{#1}\,\tilde{\Gamma}^\mu_{#1#2}\, #2}  

\newcommand{\chih}{\frac{\chi}{2}}
\newcommand{\beq}{\begin{equation}}
\newcommand{\eeq}{\end{equation}}
\newcommand{\beqa}{\begin{eqnarray}}
\newcommand{\eeqa}{\end{eqnarray}}
\newcommand{\arl}{\begin{array}{l}}
\newcommand{\earl}{\end{array}}
\newcommand{\fract}[2]{{\textstyle\frac{#1}{#2}}}
\newcommand{\fracd}[2]{{\displaystyle\frac{#1}{#2}}}
\newcommand{\thh}{\frac{\theta}{2}}
\newcommand{\oht}{\textstyle{\frac{1}{2}}}
\newcommand{\derc}[3]{\left( \frac{\partial #1}{\partial #2} \right)_{#3} }

\section{Introduction}

Neutral kaon ($\K0$) condensation treated jointly with 
$K^-$ condensation has been considered
lately by some authors \cite{Pal:2000pb,Banik:2002qu}. They  paid attention mainly to 
how the extension to the
$\K0$ condensate alters the Equation of State and
the composition of dense matter. Matter composition
 is a highly important issue for neutron star cooling
during the first few million years after the neutron star's birth.
In this period of a star's life, it is mainly cooled by neutrino emission
 from the dense core of the star.
Generally, the presence of negative boson condensate makes
matter more isospin-symmetric with the proton fraction increasing quickly with
density easily exceeding  the threshold value for the direct URCA cycle
\cite{Lattimer:1991ib}, 
which is the most effective mechanism of cooling in the dense interior
of a neutron star.
Thus, in this way kaon condensation favors fast cooling 
of neutron stars. But this conclusion should be taken carefully. It was shown
in \cite{Kubis:2002dr} that  for some class of nuclear models, 
negative kaons become so much abundant that very high proton 
fraction is required to maintain the matter neutrality. 
For very high proton fraction the direct URCA cycle is blocked again.
However, in the matter with kaon condensate, beside the direct URCA cycle,
another type of reactions is permissible.
{ It comes from the fact that the properties of nucleons in a dense 
medium change when kaon condensate is formed. Nucleons appear to be dressed
by kaons and become linear combination
of vacuum states
\cite{maxwell,Tatsumi:1988up}
\beqa
\tilde{n} &=& u\; n + v\; p\\ 
\tilde{p} &=& -\bar{v}\; n + \bar{u}\; p~,
\eeqa
where $u$ and $v$ depend on the condensate amplitude
and obey usual unitarity condition: $u\bar{u}+v\bar{v}=1$.
When condensate vanishes $\langle K \rangle \rightarrow 0$
the coefficients recover the pure nucleons, e.i. $u\rightarrow 1, v\rightarrow 0$.}
Such quasi-particles cease to be the eigenstate of charge.
Hence, besides the usual direct URCA process (dURCA) which corresponds to 
the neutron decay and its inversion, other reactions are also
possible  between nucleons dressed with kaons (kURCA):
\beq {\rm dURCA~~} \begin{array}{lll} \tilde{n}
& \leftrightarrow  & \tilde{p} + l   +~{\nu}_l \\ \end{array} 
\label{dURCA}
\eeq
\beq
{\rm kURCA~~} 
\begin{array}{lll} 
\tilde{n} & \leftrightarrow  & \tilde{n} + l  
+~{\nu}_l  \\
\tilde{p} & \leftrightarrow  & \tilde{p} + l   +~{\nu}_l~ .
\end{array}
\label{kURCA}
\eeq
For the kURCA processes it is easier to fulfill the kinematic condition 
(triangle condition) concerning the nucleon and lepton Fermi momenta.
It may be shown that independently of the value of the proton fraction,
 even for 
almost pure neutron star matter, where $x\approx 0$,
at least one of the above reactions takes place.
However, the emissivities $I$ of the three channels are not the same.
Roughly speaking, they may be classified by means of the Cabibbo angle $\th_C$
\cite{Tatsumi:1988up}:
\beq
I_{dURCA} \sim \cos^2\th_C~~~~ I_{kURCA} \sim \sin^2\th_C
\eeq
 The emissivity in the dURCA cycle is proportional to
$\cos^2\th_C$, whereas that for kURCA -- to $\sin^2\th_C$, which means that 
the kURCA branch is about two orders of magnitude less effective than the 
dURCA branch. 
In order to obtain the value of neutrino emissivity, a concrete model
of strong interaction  must be used.
As it was already mentioned there are models of $K^-$ condensation for which 
dURCA is switched off, and only less effective kURCA may cool the matter.

An interesting question is to what extent the picture changes  with 
the inclusion of the $\K0$ condensate. 
Such a component seems to be exotic but 
as it was shown by Pal et.al. in  \cite{Pal:2000pb}
that  it is quite plausible to consider 
the matter with both $K^-$ and $\K0$. 
The addition of $\K0$ presence to the model does not require additional
parameters because the form of
$\K0 N$ couplings comes from the symmetry considerations and
we need only to know the $K^- N$ coupling constant. That is important because 
the constant is not well determined quantity and we would like to
avoid any further uncertainty. The critical density for
$\K0$ condensation is always higher than that one for $K^-$. This comes from
the fact that the $\K0$ effective mass must drop to
0, whereas the $K^-$ effective mass should drop down only to the
electron chemical potential. Pal et.al. calculate the critical density 
for kaon condensation which is between $(2-3.5)n_0$ for $K^-$ and 
$(3-5)n_0$ for $\K0$, where the uncertainty comes just from the not well known
kaon-nucleon coupling strength. After $\K0$ appearance  both condensates
may coexist because such a state lowers the total energy of the system.
This also makes the Equation of State slightly softer than in the pure
$K^-$ case.
The \cite{Pal:2000pb} authors also considered several different parameterizations
of the nuclear models (with different stiffness) and showed that
 critical density for $\K0$ 
production is available in the center of a neutron star with realistic mass.
Even if the central part of the star with $\K0$ is very tiny it could
have dramatic consequences for cooling scenario. 
Page and Applegate have shown in \cite{page-appl} that even if the 
central kernel, where the direct URCA process is allowed, occupies 
only a few percent of the total mass, it cools the star in the same rate 
as it would comprise more than half of the total mass.   
One may say that if only central density of a star exceeds the threshold
value for direct URCA, the star is cooled according to the {\em fast} scenario
in which it reaches the temperature around $10^5 K$ on the scale $10^2$ years,
instead of $10^6$ years for the {\em slow} cooling driven by modified URCA processes.
So, it is important to see whether the $\K0$ presence leads to 
the fast neutrino cooling or not.
In this work, we focus on the details
of weak interactions in the matter with kaon condensate, and show
that the inclusion of $\K0$ condensation leads to
such a mixing between quasi-nucleons that removes the difference
between the dURCA and the kURCA cycles, placing them on the same footing.

{At the end of this section we wold like to refer to the issue of 
presence of hyperons in neutron star matter.
In general, the kaon condensation should be considered in common with hyperons.
Hyperons appears at lower density then the threshold for $K^-$
condensation and may move it up to higher densities even to completely block 
the production of kaons for some sets of model parameters 
\cite{Ellis:1995kz,Knorren:1995ds}. The same behavior of the threshold
density was observed in the $\K0$ case by Banik et.al. in \cite{Banik:2002qu}.
Axial coupling of kaons and hyperons opens possibility of p-wave condensation
\cite{hyp-p-wave}
and certainly affects the URCA cycles by introducing momentum dependence into
the expression for neutrino emissivity, similarly as in the pion condensation case
\cite{maxwell}. However, hyperonic star become
more and more arguable in the light of the recent observations of massive
neutron stars in X-ray binary systems \cite{x-binaries} and lately also for 
radio pulsar
\cite{Nice:2005fi} which suggest masses above 2$M_\odot$.
Different models, those based on hypernuclear observables 
\cite{Vidana:2000ew,Nishizaki:2002ih}
or those on relativistic mean field theory  
\cite{Glendenning:1991es,Knorren:1995ds}  conclude that 
maximal mass does not exceed 1.8$M_\odot$. 
One may notice that hyperons make the Equation of State much softer from 
very fundamental reasons.
Their production creates additional baryonic Fermi seas and lowers neutron
chemical potential $\mu_n$ which contributes directly to the pressure:
$P = - \ep  + \mu_n n_B$, where $\ep$ is the energy density and
$n_B$ is baryon number density.
Kaons also make the EOS softer but the scale of this effect is
model dependent and we have some freedom to avoid too soft EOS.
The kaon condensate has zero pressure
and may modify the stiffness of EOS only indirectly through effective 
masses of nucleons or lepton abundance, so various details of the
model are relevant. Nevertheless, the total effect mainly  
depends on the kaon-nucleon coupling, and previous works have shown that   
the matter with charged kaons \cite{Fujii:1995rh,Knorren:1995ds} and
with neutral kaons \cite{Pal:2000pb} is still able to support
neutron star mass above 2$M_\odot$  being in agreement with observations.
}

\section{Chiral model and weak nuclear currents}

In the context of kaon condensation, the $SU(3)_L\times SU(3)_R$ chiral model
proposed by Kaplan and Nelson \cite{Kaplan:yq} is commonly used. The current
algebra is naturally built-in into this model, so it may be used to find the
 form of hadronic currents 
needed to get the matrix elements for semileptonic reactions in the presence
of the $K^-$ and $\K0$  condensates.
The chirally symmetric part takes the following  form:

\beqa
{\mathcal L}_{\chi} & = & \frac{f^2}{4}\,\Tr \pa_\mu U \pa^\mu U^+ \,+\,
  \Tr\bar{B}(i \gamma^\mu D_\mu -  m_B) B  \nonumber  \\
 & & \,+ \, F\, \Tr \bar{B}\gamma^\mu \gamma_5 [ {\mathcal A}_\mu, B] 
  \, + \, D\, \Tr \bar{B}\gamma^\mu \gamma_5 \{ {\mathcal A}_\mu~, B\} .
\label{lan-chir}
\eeqa
Mesons are represented by the matrix
\beq
U =  \xi^2 = \exp(i\sqrt{2}\frac{M}{f})~,
\eeq
where  $M$ and $B$ include meson and baryon octet
(for notation details, see \cite{Lee:ef}):

\beq
M = \left(
  \begin{array}{ccc}
  \frac{1}{\sqrt{2}} \pi^0 + \frac{1}{\sqrt{6}} \eta_8 & \pi^+ & K^+ \\ 
  \pi^- & -\frac{1}{\sqrt{2}} \pi^0 + \frac{1}{\sqrt{6}} \eta_8 & K^0 \\
  K^- & \bar{K}^0 & -\sqrt{\frac{2}{3}} \eta_8
  \end{array}
    \right)~~~~~
B =  \left(
  \begin{array}{ccc}
  \frac{1}{\sqrt{2}} \Sigma^0 +  \frac{1}{\sqrt{6}} \Lambda & \Sigma^+ & p \\ 
  \Sigma^- & -\frac{1}{\sqrt{2}} \Sigma^0 + \frac{1}{\sqrt{6}} \Lambda & n \\
  \Xi^- & \Xi^0 & -\sqrt{\frac{2}{3}} \Lambda
  \end{array}
    \right)
\raisebox{-1em}{.}
\eeq
In order to get proper expressions for beta-type transitions for nucleons in 
the presence of $K^-$ and $\K0$, one
needs to know the conserved currents coming from the Lagrangian
(\ref{lan-chir}).
They may be found by employing the Noether theorem to the chiral transformation
$U \rightarrow LUR^+$ , $\xi \rightarrow L\xi h^+ = h \xi R^+$ ,
$B \rightarrow h B h^+$ where $L,R$ and $h$ are $SU(3)$ matrices.
It is convenient to decompose the currents into two parts:
purely mesonic and baryonic, where the latter contains baryons coupled
to meson fields:
\beq
V^\mu = V^\mu_{M} + V^\mu_B ~~~~
A^\mu = A^\mu_{M} + A^\mu_B
\label{totalsu3}
\eeq
and
\beq
V^\mu_{M,a} = -i \frac{f^2}{4} \Tr \lambda_a (U^+\pa^\mu U + U\pa^\mu U^+)
\eeq
\beq
A^\mu_{M,a} = i \frac{f^2}{4} \Tr \lambda_a (U^+\pa^\mu U - U\pa^\mu U^+)
\eeq
\beq
V^\mu_{B,a} = \,\frac{1}{4} \Tr \bar{B}\gamma^\mu [u_+^a,B] \,+\, 
\frac{F}{4}\Tr \bar{B}\gamma^\mu \gamma_5 [u_-^a,B] \,+\,
\frac{D}{4}\Tr \bar{B}\gamma^\mu \gamma_5 \{u_-^a,B\}
\label{vsu3}
\eeq
\beq
A^\mu_{B,a} = \,\frac{1}{4} \Tr \bar{B}\gamma^\mu [u_-^a,B] \, + \,
\frac{F}{4}\Tr \bar{B}\gamma^\mu \gamma_5 [u_+^a,B] \,+\,
\frac{D}{4}\Tr \bar{B}\gamma^\mu \gamma_5 \{u_+^a,B\}
\label{asu3}
\eeq  
where $u_\pm^a = \xi^+ \lambda_a \xi \pm \xi \lambda_a \xi^+$.
The expression for axial mesonic current $A_M^\mu$ allows one to identify
the parameter $f$ with the pion decay constant $f_\pi$, whereas the 
baryonic part is relevant for semileptonic decays of baryons.
The above formulae are similar in their form to the weak nuclear 
currents presented
in \cite{Serot:2004rc} for the $SU(2)\times SU(2)$ chiral model.

The ground state of
matter with kaon condensate is described by Fermi seas of baryons 
and the non-vanishing expectation value of kaon fields.
According to the Baym theorem \cite{baym}, the kaon mean field
acquires time dependence
\beq
\langle K^\pm \rangle = \frac{f\th}{\sqrt{2}} \exp(\pm i \mu t)~~~
\langle K^0 \rangle = \langle \K0 \rangle = \frac{f\phi}{\sqrt{2}}
\label{kmean}
\eeq
recalling that for neutral kaons, their mean field is independent of time, as
the chemical potential for  neutral particles vanishes. 
The quantities $\th, \phi$ are non-dimensional condensate amplitudes,
useful to parameterize the condensate state.
These condensate amplitudes acquire the meaning of rotation angles in
chiral space, if, instead of the matrix
element $\langle\tilde{p},\th,\phi|{\mathcal O}|\tilde{n},\th,\phi\rangle$
of any operator $\mathcal O$ between quasi-nucleons,  one  considers  
$\langle{p}|{\mathcal U}^+(\th,\phi)\,{\mathcal O}\,{\mathcal U}(\th,\phi)|n\rangle$
-- the matrix element between normal nucleons, but with a rotated $\mathcal O$.
In this way,
for example, the isospin-raising operator $V_{1+i2}$,  relevant 
for the beta decay, in the condensate state 
becomes linear combinations of all currents from the octet $V_a$, leading
to transitions forbidden in the normal state, as reactions shown in (\ref{kURCA}). 
{ In our approach we do not need to
know the explicit form of the $\cal U(\th,\phi)$  because any 
nuclear current may be find directly by putting expectations 
values of kaon fields (\ref{kmean}) into equations (\ref{vsu3},\ref{asu3}).
In this derivation, the important quantity  is the kaon field
matrix $\xi$, which now takes the form
\beq
\xi = \frac{1}{\chi^2}\left(
\begin{array}{ccc}
{\phi }^2 + {\theta }^2\cos\fracd{\chi}{2} &
 -2\,\ee^{i t\mu }\theta\phi\,\sin^2 \fracd{\chi }{4} &
 i \ee^{i t\mu }\theta \chi \sin \fracd{\chi }{2} \\
 -2\,\ee^{-it\mu }\theta \phi\,{\sin^2 \fracd{\chi }{4}} &
 {\theta }^2 + {\phi}^2\cos \fracd{\chi }{2} & 
i \phi \chi \sin \fracd{\chi }{2} \\
i \ee^{-it\mu } \theta \chi \sin \fracd{\chi }{2}&
 i \phi \chi \sin \fracd{\chi }{2} &
{\chi }^2\cos \fracd{\chi }{2}~~ 
\end{array}
\right),
\label{ximean}
\eeq
where $\chi^2=\th^2+\phi^2$.
The weak hadronic current $J_\mu$ has the usual $V\!-\!A$ structure, but 
for the strange particle case, besides the isospin-raising part 
$V_{1+i2} \!-\! A_{1+i2}$, one must include the strangeness-changing
part coming from the $SU(3)$ octet. Following the Cabibbo theory,
the hadronic current is
\beq
J_\mu = \cos\th_C (V_{1+i2} - A_{1+i2})_\mu + \sin\th_C (V_{4+i5} - A_{4+i5})_\mu .
\label{weakhadr}
\eeq
The rate of the reactions which operate in the URCA cycles is described by the 
baryonic part $V_B, A_B$ of the total $SU(3)$ current (\ref{totalsu3}).
Knowing the form of the kaon field matrix $\xi$,
 we may calculate the full octet of currents. 
Putting them into (\ref{weakhadr}), one
gets the weak hadronic current $J^\mu$ in the presence of 
the $K^-$ and $\K0$ condensate
\beq
\begin{split}
J^\mu = &\;\frac{\cos\th_C}{\chi^4}\bigl\{ 
   \gammc{p}{n}\, \bigl[(\th^4+\phi^4)\cos\chih + 
        \frac{\th^2\phi^2}{2}(3+\cos\chih) \bigl]\\
& \phantom{\frac{\cos\th_C}{\chi^4}\bigl\{} 
+\; (\gammc{n}{n}+\gammc{p}{p}) \, 2\th\phi \:\ee^{-i\mu t} \sin^2\frac{\chi}{4}
    \\
& \phantom{\frac{\cos\th_C}{\chi^4}\bigl\{} 
+\;\gammc{n}{p} \,\bigl( 2\th\phi \:\ee^{-i\mu t} \sin^2\frac{\chi}{4}\bigr)^2
  \bigr\}\\
&\;\frac{\sin\th_C}{\chi^3}\sin\chih\bigl\{  
\gamms{p}{n}\,  (\phi^2 + \th^2\cos\chih) i \phi \\
& \phantom{\frac{\sin\th_C}{\chi^3}\sin\chih\bigl\{}
+\;  (\gamms{n}{n}+\gamms{p}{p}) \, i\th\:\ee^{-i\mu t} \\
& \phantom{\frac{\sin\th_C}{\chi^3}\sin\chih\bigl\{}
+\;    \gamms{n}{p}\, 2 i \ee^{-2 i\mu t} \th^2 \phi \sin^2\frac{\chi}{4}\bigr\},
\end{split} 
\label{weakK}
\eeq
where  the matrices $\Gamma_{ij}$ and $\tilde{\Gamma}_{ij}$ are linear 
combinations of the Dirac matrices, explicitly given  in the Appendix.
}

\section{Neutrino emissivity}

At this point, we are ready to derive the transition rate for
beta processes in  matter with the $K^-, \K0$ condensate. 
Because the energy of nucleons and leptons is much smaller than the 
mass of $W^\pm$ particles, it is sufficient to use the Fermi theory of
weak interactions, for which the Hamiltonian takes the usual form:
\beq
H_{weak} = \frac{G_F}{\sqrt{2}} J_\mu l^\mu.
\eeq
Where $J^\mu$ is hadronic and $l^\mu$ is leptonic weak current.
In the case of charged-kaon ($K^-$) condensate only, $J^\mu$ derived in the 
previous section reduces to simpler form
\beq
\begin{split}
J^\mu_{\phi=0}& = \cos\th_C \,\bar{p}\gamma^\mu (1-g_A \gamma_5) n \cos\thh \\
&\; + \;
\sin\th_C\, [\bar{n}\gamma^\mu (1+\Delta g\,\gamma_5)n \, + 
             \, 2\bar{p}\gamma^\mu (1\!-\!F\gamma_5)p]
\frac{i}{2} \sin\th
\end{split}
\label{weakth}
\eeq
where $g_A\! =\! D+F$ and $\Delta g\!=\!D-F$, and which is equivalent
 to the results already presented in \cite{Fujii:1993cf}.
Comparing (\ref{weakth}) and (\ref{weakK}), one may note that the extension 
to neutral kaons introduces an additional
term $\sim \bar{n}\gamma^\mu p$, which is absent in the pure $K^-$
condensate case. This term opens a new URCA channel, let us call it k0URCA:
\beq
{\rm k0URCA}~~~\tilde{p} \leftrightarrow \tilde{n} + e + \nu_e
\label{k0URCA}
-\eeq
Although slightly exotic, this kind of "proton decay" is possible in 
dense matter, as one must remember that quasi-nucleons represent mixed states
of normal nucleons and do not possesses a well-determined charge. Moreover, the
kURCA transitions, i.e. transitions between the same type of quasi-nucleons
 (\ref{kURCA}), are
now also present in the  $cos\th_C$-dependent part of the weak current. 
This means that for the
dense matter state, where the two kinds of condensates are simultaneously present
($\phi\ne0$, $\th\ne 0$) , both  the kURCA and dURCA cycles will take place at the
same  rate. This is an important result as it is easier to fulfill 
the triangle condition for the Fermi momenta of nucleons in the case of 
the kURCA channel.

The energy per unit volume and time released in one cycle due to the neutrino 
emission is equal to the product of the neutrino energy and the
doubled beta decay rate
for a given quasi-particle $i=\tilde{p}$ or $\tilde{n}$
\beq
I_{URCA} = \frac{2}{(2\pi)^{12}} \int d^3p_id^3p_fd^3p_ed^3p_\nu
 f_i(1-f_f)(1-f_e)
W_{if} \ep_\nu .
\label{Iurca}
\eeq
The decay rate $W_{ij}$ for a transition:  $i\rightarrow j+e+\nu_e$ is
given by the expression
\beq
W_{if} = (2 \pi)^4 \delta(\ep_f-\ep_i-\ep_e-\ep_\nu)
 \delta({\pmb p_f-\pmb p_i-\pmb p_e-\pmb p_\nu})
 |\langle f\,e\,\nu_e \,|\,H_{weak}\,|\,i\,\rangle|^2
\label{Wif}
\eeq
The squared matrix element may be factorised in a standard manner
\beq
|\langle f\,e\,\nu_e \,|\,H_{weak}\,|\,i\,\rangle|^2=H^{\mu\nu}L_{\mu\nu}
\label{matrelem}
\eeq
where the leptonic tensor is 
\beq
L^{\mu\nu} =\frac{1}{\ep_e\ep_\nu} (p_e^\mu p_{\anu}^\nu + p_e^\nu p_{\anu}^\mu
 - p_e\!\cdot\! p_{\anu}\,g^{\mu\nu} +
 i \epsilon^{\mu\nu\rho\sigma} p_{e\rho} p_{\anu\sigma} )
\eeq
and the hadronic tensor includes the hadronic weak current $J^\mu$
with summation over the nucleon spin states
\beq
H^{\mu\nu} = \sum_{s,\sp} \langle p_f,\sp|\,J^\mu\,|p_i,s\rangle
\langle p_i,s|\,J^{+\nu}\,|p_f,\sp\rangle .
\eeq
For non-relativistic nucleons, the hadronic tensor becomes momentum-independent 
$H^{\mu\nu}= 2 (|v|^2\delta_{0\mu}\delta_{0\nu} + 
|a|^2\delta_{i\mu}\delta_{i\nu})$, where $a,v$  are the axial and vector
parts of the hadronic current (\ref{weakK}). The neutrino momentum 
$|{\pmb p_\nu}|$ is of the order of thermal energy $T$, which means it is much
smaller than the momenta of nucleons and that of electron, 
and may be neglected in the delta function in (\ref{Wif}).
Finally, the matrix element (\ref{matrelem}) may be treated as a constant, 
being independent of particle momenta.
The phase space in (\ref{Iurca}) is  weighted only by
the three Fermi-Dirac distribution functions $f_i$ for $\tilde{p},\tilde{n},e$,
because the neutrino leaves the matter freely.
In order to calculate the emissivity  integral, one may use 
the so-called {\em phase-space decomposition}, a very useful technique to obtain 
the reaction rate for strongly-degenerated systems 
\cite{Shapiro:1983du,Yakovlev:2000jp}.
After this calculation, neutrino emissivity may be finally written down 
as  the following expression:
\beq
I_{URCA}=\frac{457\,\pi}{20\,160} T^6 m^*_p m^*_n m^*_e |M_{ij}|^2 \Theta_{ife}
\eeq
where $\Theta_{ife}$ is a step function corresponding to the triangle condition:
it is equal to 1 if vectors: ${\pmb p_f,\pmb p_i,\pmb p_e}$ form a closed 
triangle, and 0 otherwise. 
$M_{ij}$ is the momentum-independent matrix element
(\ref{matrelem}) corresponding to different types of reactions, and
$m^*_i$ are effective masses of nucleons and electron. 
In the case of the $K^-$ and $\K0$ condensation, 
there are four channels for different URCA cycles: one in (\ref{dURCA}) two in 
(\ref{kURCA}) and one in (\ref{k0URCA}).
The last one is typical for the presence of neutral kaons, whereas
the reactions (\ref{kURCA}) are connected with  the charged kaon condensate.
Of course, the first one, i.e. the dURCA cycle, is possible in normal $npe$
 matter as well as in matter with kaons.
All of them belong to the class of {\em direct} URCA processes, 
where three
degenerated fermions only take part in the cycle, which 
can be seen in the temperature
dependence $\sim T^6$. So, the four different channels are only distinguished
by the means of their matrix elements $|M_{ij}|^2$, which now depend on the 
amplitudes of the two condensates  $\th$ and $\phi$:
\begin{align}
\begin{split}
{\rm dURCA}~~~~|M_{np}|^2 =&
 \;2\,G_F^2\,( 1 + \,3\, g_A^2)
  \frac{\left( {\phi }^2 + {\theta }^2\cos\fracd{\chi }{2} \right)^2}{\chi^8}\times \\
&  \Big[ {\left( {\theta }^2 + {\phi }^2\cos\fracd{\chi }{2}  \right) }^2
       {{{\cos^2\theta }_C}} \, +  \,
    \phi^2\chi^2\,{\sin^2\fracd{\chi}{2}}\;{{{\sin^2\theta}_C}}
      \Big]
\label{mdurca0} 
\end{split}\\
\begin{split}
{\rm kURCA}~~~~|M_{nn}|^2 =&
 \;8\,G_F^2\,\cos^2\th_C\frac{\th^2\phi^2\sin^4\frac{\chi}{4}}{\chi^8}
\Big\{ \Big[ \th^2+\chi^2 + (\phi^2+\chi^2)\coschh\Big]^2\\
& \qquad + 3\Big[2\,\dga\chi^2 \cos^2\frac{\chi}{4} - 
    g_A\left(\th^2+\phi^2\coschh\right) \Big]^2\Big\}\\
& +\;2\,G_F^2\,\sin^2\th_C\frac{\th^2\sin^2\frac{\chi}{2}}{\chi^6}
\Big\{ \Big[ \phi^2 - (\phi^2+\chi^2)\coschh\Big]^2\\
& \qquad\qquad + 3\Big[ \dga\chi^2\coschh - 
    2 g_A\phi^2\sin^2\frac{\chi}{4}\Big]^2\Big\}
\label{mknurca0}
\end{split}\\
\displaybreak[0]
\begin{split}
{\rm kURCA}~~~~|M_{pp}|^2 =&
 \;\,G_F^2\,\cos^2\th_C\frac{\th^2\phi^2\sin^4\frac{\chi}{4}}{\chi^8}
\Big\{ \Big[ \phi^2+\chi^2 + (\th^2+\chi^2)\coschh\Big]^2\\
& \qquad + 3\Big[ 2\,\dga\chi^2\cos^2\frac{\chi}{4} - 
    g_A\left(\phi^2+\th^2\coschh\right) \Big]^2\Big\}\\
& +\;2\,G_F^2\,\sin^2\th_C\frac{\th^2\sin^2\frac{\chi}{2}}{\chi^6}
\Big\{ \Big[ \phi^2 + (\th^2+\chi^2)\coschh\Big]^2\\
& \qquad\qquad + 3\Big[ \dga\chi^2\coschh - 
    g_A\left(\phi^2+\th^2\coschh\right) \Big]^2\Big\}
\label{mkpurca0}
\end{split}\\
\begin{split}
{\rm k0URCA}~~~~|M_{pn}|^2 =&
\;32\,G_F^2\,( 1 + \,3\,g_A^2 )
\frac{\th^4 \phi^2\, \sin^6{\fracd{\chi}{2}}}{\chi^8}\times  \\
& \left(\phi^2 \sin^2\fracd{\chi}{4}\cos^2\th_C \,+
\, \chi^2 \cos^2\fracd{\chi}{4}\sin^2\th_C \right)
\end{split}
\end{align}
As was already shown in \cite{Fujii:1993cf}, for the case of the $K^-$
condensate only ($\phi\rightarrow 0$) these matrix elements takes a much 
simpler form
\begin{align}
|M_{np}|^2& = 2\,G_F^2 \,(1+3\,g_A^2)\cos^2\th_C  \cos^2\thh   
\label{mdurca} \\
|M_{nn}|^2& = \fract{1}{2}G_F^2\,(1+3\,\dga^2) \sin^2\th_C \sin^2\th
\label{mknurca} \\
|M_{pp}|^2& = 2\,G_F^2\,(1+3\,F^2) \sin^2\th_C  \sin^2\th 
\label{mkpurca} \\
|M_{pn}|^2&= 0
\label{mk0urca}
\end{align}
where, for the transition $\tilde{p}\rightarrow\tilde{n}+e+\bar{\nu}$
which  is not possible in the pure $K^-$ 
condensate, the corresponding matrix element vanishes.
By comparing expressions (\ref{mknurca0},\ref{mkpurca0}) and
(\ref{mknurca},\ref{mkpurca}), one may see how the inclusion of $\K0$
introduces the terms proportional to $\cos^2\th_C$
and, in this way, makes the kaon-induced URCA processes
at the same level of intensity as the direct URCA cycle in normal 
matter. This was already noted in the discussion of the weak hadronic current
(\ref{weakK}). 
\begin{figure}[t]
\includegraphics[width=.8\textwidth]{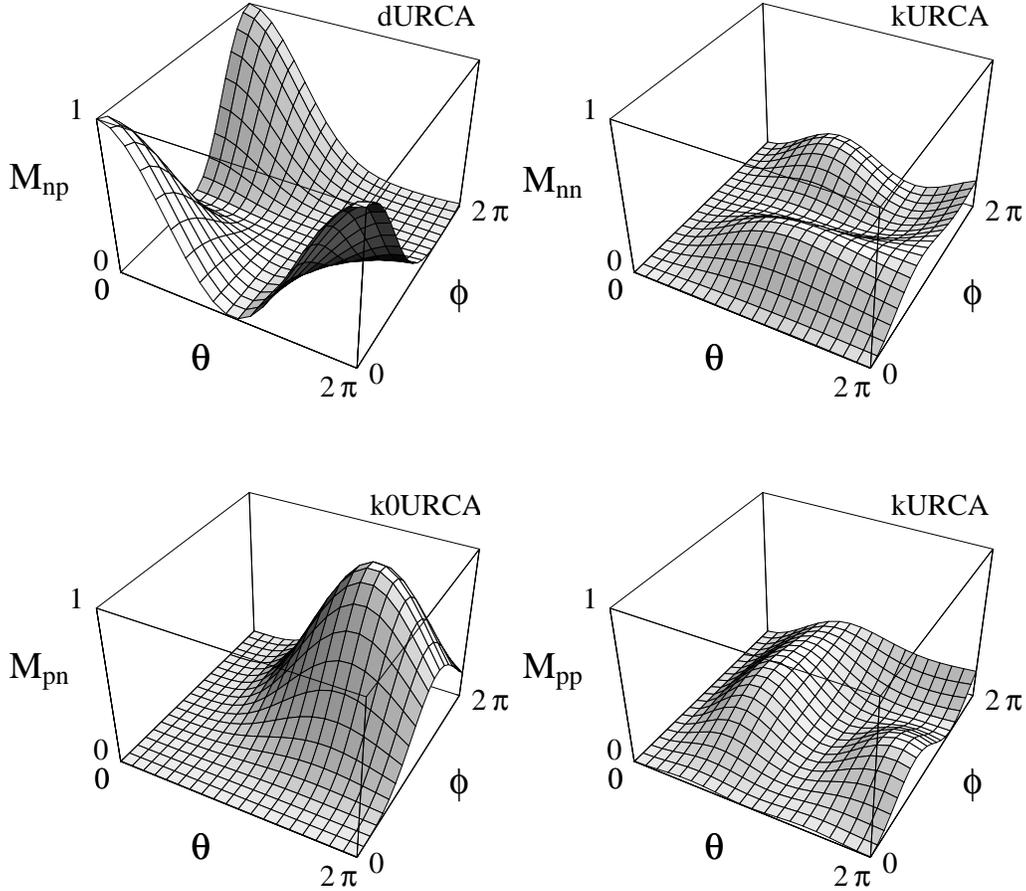}
\caption{Normalized matrix element for different URCA channels 
as a function of condensate amplitudes $\th$ and $\phi$.}
\label{figmif}
\end{figure}
The matrix elements $M_{if}$ are also 
highly $\th$- and $\phi$-dependent. This dependence is shown in Fig.\ref{figmif},
where the values of matrix elements were normalized to the maximal value for the
dURCA cycle, which is $M_{np}^{max} = 2 G_F^2 (1+3 g_A^2)$.
The behavior of $\th$ and $\phi$
with matter density depends on the model of strong interactions used for
dense matter description \cite{Kubis:2002dr,Pons:2000iy}.
Both the unknown strength
of the kaon-nucleon coupling and the details of interactions in the 
non-strange sector
affect the condensate behavior. However, roughly speaking, these papers
showed that the typical  value of the $K^-$ condensate  amplitude  $\th$ is
around $1$, and in some cases it may be somewhat greater than $\pi/2$.
Therefore, one may suspect that the amplitude of the $\K0$ condensate
takes similar values.
A careful look at the plots in Fig.\ref{figmif} shows that different cycles 
reach their maxima in different regions of the $\th-\phi$ plane, so
one may conclude that almost independently of the concrete values of 
the $K^-$ and $\K0$ 
amplitudes, there always exists one cycle with emissivity approximately 
equal to the maximal value of $M_{if}$, i.e. $2 G_F^2 (1+3 g_A^2)$.
The given cycle works in neutron star matter when  the corresponding 
triangle condition is satisfied:

\begin{align}
\rm dURCA~~and~~k0URCA~~~~& |k_p - k_n| < k_e < k_p+k_n \\
\rm kURCA~for~\tilde{n}~~~~& 2 k_n > k_e \\
\rm kURCA~for~\tilde{p}~~~~& 2 k_p > k_e
\end{align}
The above inequalities show that independently of the proton and lepton
 abundances
at least one of the URCA channel is opened. Thus, the final conclusion is 
that the simultaneous presence of the $K^-$ and $\K0$ condensate leads to matter
cooled very fast with intensity of the order of the fastest direct URCA
cycle.

\section*{Conclusions}
The extension of the charged kaon to the neutral kaon condensate leads to such
 mixing
between  the components of currents from the $SU(3)$ octet that two new
features emerge. First, the $\K0$ condensate opens an 
additional channel for the URCA process (quasi-proton decay). Second,
the $\K0$ presence results in emissivity for the kaon-induced URCA processes
scaling no longer with $\sin^2\th_C$ but obtaining a contribution 
that scales as $\cos^2\th_C$. This puts the kaon-induced URCA
 at the same level of importance as the normal
direct URCA cycle. 
Moreover, the triangle condition says that different cycles
are opened in different 
matter compositions, covering the whole range froqqm pure neutron to
pure proton matter, ($0<x<1$). Therefore, finally one may conclude that
matter with the $K^-$ and $\K0$ condensate is cooled at the level of
the most effective (direct) URCA cycle - regardless of its detailed composition.

\section*{Acknowledgments}
This work was partially supported by the Polish State Committee for
Scientific Research Grant 2P03D 02025.

\section*{Appendix}
\appendix 

Below, there are matrices $\Gamma_{ij}$ and $\tilde{\Gamma}_{ij}$ 
that appear in the expression
for the  hadronic current (\ref{weakK}), for the $\cos\th_C$-dependent part:
\[
\begin{split}
\Gamma^\mu_{pn}& = \gamma^\mu(1-g_A\gamma_5)\\
\Gamma^\mu_{pp}& = \gamma^\mu\bigl[-(\phi^2\!+\!\chi^2)-(\th^2\!+\!\chi^2)\cos\chih \\
 & \phantom{\gamma^\mu\bigl[ +} + 
 \gamma_5\bigl(g_A(\phi^2\!+\!\th^2\cos\chih)-2\,\Delta g\,\chi^2\cos^2\frac{\chi}{4}
  \bigr)\bigr]\\
\Gamma^\mu_{pp}& = \Gamma^\mu_{nn}(\phi\leftrightarrow\th)\\
\Gamma^\mu_{np}& = \gamma^\mu(1+g_A\gamma_5)
\end{split}
\]
and for the $\sin\th_C$-dependent part:
\[
\begin{split}
\tilde\Gamma^\mu_{pn}& = -\gamma^\mu(1-g_A\gamma_5)\\
\tilde\Gamma^\mu_{pp}& = \gamma^\mu\bigl[-2\phi^2\cos^2\frac{\chi}{4} +  
 \gamma_5\bigl(g_A(\phi^2\!+\!\th^2\cos\chih)-\Delta g\,\chi^2\cos\frac{\chi}{2}
  \bigr)\bigr]\\
\tilde\Gamma^\mu_{nn}& = \gamma^\mu\bigl[-\phi^2+(\phi^2\!+\!\chi^2)\cos\chih + 
\gamma_5(2g_A\phi^2\sin^2\frac{\chi}{4} + 
               \Delta g\,\chi^2\cos\frac{\chi}{2}\bigr]\\
\tilde\Gamma^\mu_{np}& = \gamma^\mu(1-g_A\gamma_5) ~.
\end{split}
\]

\bibliography{k0emis}

\end{document}